\documentclass[%
 reprint,
 amsmath,amssymb,
 aps,apj
]{emulateapj}

\usepackage{graphicx}
\usepackage{dcolumn}
\usepackage{bm}
\usepackage{graphicx,latexsym,color,dsfont,float}
\usepackage{amsmath,amsfonts,amsbsy,amssymb,amsthm}
\usepackage[utf8]{inputenc}
\usepackage[mathscr]{euscript}
\usepackage[T1]{fontenc}
\usepackage{mathtools,physics}
\usepackage[english]{babel}
\usepackage{scalerel}
\usepackage{float}
\usepackage[medium]{titlesec}
\usepackage{booktabs} 
\usepackage{multirow}
\usepackage{bm}
\usepackage{subfigure}
\usepackage[bookmarksnumbered=true,bookmarksopen=true]{hyperref}
\hypersetup{
	colorlinks=true,
	linkcolor=blue,
	filecolor=blue,      
	urlcolor=blue,
	citecolor=blue
}

\begin{document}


\title{Gravitational Collider Physics via Pulsar-Black Hole Binaries II: \\Fine and Hyperfine Structures are Favored}

\author{Xi Tong$^{1,2}$}
\email{xtongac@connect.ust.hk}
\author{Yi Wang$^{1,2}$}
\email{phyw@ust.hk}
\author{Hui-Yu Zhu$^{1,2}$}
\email{hzhuav@connect.ust.hk}
\affiliation{${}^1$Department of Physics, The Hong Kong University of Science and Technology, \\
	Clear Water Bay, Kowloon, Hong Kong, P.R.China}
\affiliation{${}^2$The HKUST Jockey Club Institute for Advanced Study, The Hong Kong University of Science and Technology, \\
	Clear Water Bay, Kowloon, Hong Kong, P.R.China}

\begin{abstract}
A rotating black hole can be clouded by light bosons via superradiance, and thus acquire an atom-like structure. If such a gravitational atom system is companioned with a pulsar, the pulsar can trigger transitions between energy levels of the gravitational atom, and these transitions can be detected by pulsar timing. We show that in such pulsar-black hole systems, fine and hyperfine structure transitions are more likely to be probed than the Bohr transition. Also, the calculation of these fine and hyperfine structure transitions are under better analytic control. Thus, these fine and hyperfine structure transitions are more ideal probes in the search for gravitational collider signals in pulsar-black hole systems.
\end{abstract}

\maketitle


\section{Introduction}

Compact objects play an important role in astrophysics and particle cosmology. With their astronomical mass squeezed into a compact region of spacetime, these objects provide an ideal platform for high energy astrophysical phenomena as well as a test ground for gravity \citep{Shapiro:1983du,2007coaw.book.....C}.

Arguably, the most notable of the compact objects are pulsars and black holes. Pulsars are usually rotating neutron stars that emit beams of electromagnetic radiation from their magnetic poles. The beam sweeps across the earth periodically due to the rotation. To a distant observer, despite its complex internal dynamics, a pulsar can be simply viewed as a clock that ticks by emitting periodic radio pulses. The precise periodicity of the pulsar makes it an excellent timing tool. For instance, pulsar timing is used to measure the orbital decay of binary systems \citep{Hulse,Hulse2}, to detect low-frequency Gravitational Waves (GWs) \citep{Hobbs:2009yy,SKA,NANOGrav}, to probe gravity in the strong-field regime \citep{1007.0007}, and even to serve as autonomous space navigation beacons \citep{1305.4842}.

Black holes also play a central role in modern physics. Although isolated stationary black holes are classically characterized only by three parameters, they are known to carry more structures in the presence of perturbations \citep{Konoplya:2011qq,Hawking:2016msc}. Not only does the horizon emit Hawking radiation quantum mechanically \citep{Hawking:1974rv}, which inspired numerous studies on the long-standing information paradox \citep{Hawking:1974sw}, a rotating black hole carries a dissipative ergoregion capable of radiating particles on a classical level \citep{1971JETPL..14..180Z,Press:1972zz}. This phenomenon known as superradiance has also been widely studied for over half a century (see a comprehensive review given by \citet{Brito:2015oca}). For bosonic particles with mass $\mu\lesssim(G M_B)^{-1}$, where $M_B$ is the mass of the black hole, superradiance triggers an instability in the spectrum of the black hole bound states \citep{Damour:1976kh}. This leads to the formation of a bosonic cloud of size $r_1\sim\mathcal{O}(10^{1}\text{-}10^3) GM_B$ around the black hole, with an energy spectrum similar to that of the hydrogen atom. In isolation, such a gravitational atom emits monochromatic GWs through pair annihilations as well as spontaneous level transitions \citep{Arvanitaki:2010sy}. When a binary companion is introduced, the periodic orbital motion may hit the resonance band of the gravitational atom and induce Landau-Zener transitions \citep{10011873546,Zener:1932ws} between different energy levels. The backreaction effect produces floating or sinking/kicked orbits observable from the GW signatures emitted by the binary. This recently proposed framework aimed at probing ultralight bosons is known as Gravitational Collider Physics (GCP) \citep{baumann2019probing,baumann2019spectra,baumann2020gravitational}.

However, GW \citep{Ng:2020jqd} is not the only observation channel for GCP resonances and ultralight bosons. The backreaction on the binary can be naturally viewed as a change of orbital period derivative, $i.e.$, a timing problem. Given the simple yet accurate time periodicity of the pulsar, it is natural to consider the Pulsar-Black Hole (PSR-BH) binary as a viable probe of the GCP resonances. This PSR-BH-radio observation channel has recently been verified in \citep{oringin} for Bohr transitions of the gravitational atom\footnote{See other PSR-BH approaches to ultralight bosons in \citep{4,5}.}. 

Bohr transitions change the principal quantum number, thus due to the narrow mass range of the pulsar, the corresponding resonance frequency is relatively high, and the orbital period can be as short as $P_r\sim \mathcal{O}(1)$s. Hence the pulsar timing accuracy proves to be always sufficient. However, there are still several problems faced by the PSR-BH-radio channel for Bohr transitions. (i) The short binary period during Bohr transitions suggests that the binary is near the end of the inspiral process. Binaries with such a short period are statistically disfavored. Given the fact that the number of observable PSR-BH binaries in our Galaxy is limited \citep{Faucher_Gigu_re_2011,Shao:2018qpt,Chattopadhyay:2020lff}, the event rate for Bohr transitions in the PSR-BH binaries may be extremely low. (ii) During a Bohr transition, the binary separation is comparable to the size of the boson cloud. This may threaten the validity of the quadrupole approximation, leading to the inadequacy of considering a narrow resonance with $\Delta m=2$ only. Additional effects such as dynamical friction \citep{1907.13582}, upscattering effects \citep{Wong:2020qom}, and the emergence of additional molecular states \citep{2010.00008} may also dramatically change the prediction.

Compared to the Bohr transition, fine/hyperfine GCP transitions are observationally more probable and theoretically cleaner to analyze. In this work, we set out to analyze the fine/hyperfine GCP transitions of PSR-BH binaries for ultralight scalar bosons. The advantages of probing GCP with fine/hyperfine transition are: (i) The fine and hyperfine splittings are suppressed by extra factors of $\alpha^2$ and $\alpha^3$ respectively, where $\alpha\equiv GM_B\mu\ll1$ is the gravitational fine structure constant. Therefore, they have a much longer resonance period. In addition, some fine/hyperfine transitions give floating orbits which enjoy an extremely long duration. This dramatically increases the event rate. (ii) The increase in resonance period is accompanied by the increase in the binary separation, which is now much greater than the cloud radius. This ensures the validity of the quadrupole approximation and the narrow resonance.

Apparently, a disadvantage of fine/hyperfine transitions is that the signal is weaker than Bohr transitions. The long orbital period gives rise to an orbital decay that may be too tiny to be detected by the first generation of space GW detectors \citep{LISA,Taiji,tianqin,decigo}. However, thanks to the well-established timing accuracy of pulsars, a long-term observation is sufficient to capture the fine/hyperfine resonances, as we will show below.

This paper is organized as follows. In Sect.~\ref{GCPviaPSRBHReview}, we review some technical details of the gravitational atom and GCP transitions. In Sect.~\ref{BohrVsFine}, we discuss the problems faced by the Bohr transitions and motivate our study for fine/hyperfine transitions. Then in Sect.~\ref{FineHyperfine}, we focus on the major fine/hyperfine transitions induced by the quadrupole moment of the orbiting pulsar and analyze their observational feasibility. We conclude and give future prospects in Sect.~\ref{Conclusion}.

\section{The Gravitational Atom}\label{GCPviaPSRBHReview}
In this section, we briefly review the basic aspects of the gravitational atom and GCP following \citet{baumann2019probing} and \citet{baumann2020gravitational}. A Kerr black hole is equipped with a dissipative ergosphere that can amplify incoming waves \citep{Penrose:1969pc,1971JETPL..14..180Z}. For a massive bosonic field, its mass serves as a natural mirror that reflects back the amplified modes \citep{Press:1972zz,Cardoso:2004nk}, from which an instability is generated \citep{Damour:1976kh}. This superradiance instability leads to the growth of a bosonic cloud, whose behavior is governed by the Schrödinger equation with corrections from Kerr spacetime, 
\begin{equation}
i\partial_t\psi(t,\vec{x})=\left(-\frac{1}{2\mu}\partial_{\vec{x}}^2-\frac{\alpha}{r}+\mathcal{O}(\alpha^2)\right)\psi(t,\vec{x})~,
\end{equation}
where $\alpha\ll 1$ guarantees the validity of non-relativistic expansion. The solutions of the Schrödinger equation with in-going boundary condition at the horizon are atomic states $|nlm\rangle$ labeled by the principal, angular and magnetic quantum numbers\footnote{Throughout this paper, we assume the ultralight boson is a (pseudo)scalar.}. The frequency of each eigenstate is in general complex: $\omega_{nlm}=E_{nlm}+i\Gamma_{nlm}$, with
\begin{equation}
\begin{aligned}
  E_{nlm}&=\mu\Bigg(1-\frac{\alpha^2}{2n^2}-\frac{\alpha^4}{8n^4}-\frac{(3n-2l-1)\alpha^4}{n^4(l+1/2)}\\
  &~~~~~~~~~+\frac{2\tilde{a}m\alpha^5}{n^3l(l+1/2)(l+1)}+O(\alpha^6)\Bigg), \label{E}
\end{aligned}
\end{equation}
and
\begin{equation}
  \Gamma_{nlm}=2\tilde{r}_+C_{nlm}(m\Omega_H-\omega_{nlm})\alpha^{4l+5}~,\label{Gamma_nlm_Detweiler}
\end{equation}
where $\tilde{r}\equiv r/M_B$ and $\tilde{a}\equiv a/M_B\lesssim 1$ is the dimensionless black hole spin. For simplicity, we will assume $\tilde{a}\simeq 1$ throughout this work, since astrophysical BHs in binaries generally have a large spin \citep{OShaughnessy:2005ias,Nielsen:2016kyw}. $C_{nlm}$ can be found in \citet{baumann2019probing}. Notice that although (\ref{Gamma_nlm_Detweiler}) is derived under Detweiler’s approximation \citep{Detweiler:1980uk} with $\alpha\ll 1$, numerical studies have confirmed it validity for $\alpha<0.5$ \citep{Brito:2015oca}. For a positive $\Gamma_{nlm}$, the total mass of the state $M_{nlm}$ will grow at a timescale $T^{\text{(growth)}}_{nlm}\equiv\Gamma_{nlm}^{-1}$ until saturation, then it slowly depletes via the emission of GWs. The depletion power of a highest helicity state with $l=m$ is
\begin{equation}
    \dot M_{nlm}=-B_{nl}\left(\frac{M_{nlm}}{M_B}\right)^2 \alpha^{4l+10}~.\label{depleteApprox1}
\end{equation}
The coefficient $B_{nl}$ is computed by \citet{Bnl} in the flat spacetime limit as
\begin{equation}
    B_{nl}=\frac{16^{l+1}l(2l-1)\Gamma(2l-1)^2\Gamma(l+n+1)^2}{n^{4l+8}(l+1)\Gamma(l+1)^4\Gamma(4l+3)\Gamma(n-l)^2}.\label{depleteApprox2}
\end{equation}
However, we caution the reader that numerical analysis shows that (\ref{depleteApprox2}) underestimates the depletion power by approximately one order of magnitude \citep{Bnl,Brito:2014wla}. The scaling powers of $M_{nlm}$ and $\alpha$, in contrast, are robust for $\alpha\ll 1$. Nevertheless, due to its analytical generality for all $n$ and $l$, we will adopt (\ref{depleteApprox2}) to estimate the depletion time scale,
\begin{equation}
	T_{nlm}^{\text{(deplete)}}\approx B_{nl}^{-1}\frac{M_B^2}{M_{nlm,0}}\alpha^{-4l-10}~,\label{DepleteTimeApprox}
\end{equation}
and warn the reader about the potential $\mathcal{O}(10)$ uncertainty due to (\ref{depleteApprox2}). Here $M_{nlm,0}$ is the initial mass of the cloud state, whose value at saturation can be estimated using angular momentum conservation (see Table.~1 in \citet{baumann2020gravitational}).

Now let us come to the binary system case. The motion of a binary companion will generate a periodic tidal perturbation on the gravitational atom, leading to level crossings in the atomic spectrum. At a large binary separation, the tidal perturbation is solely dependent on the mass of the binary companion. Thus it can be any astronomical object compact enough to fit into the orbit. In particular, we consider a pulsar of mass $M_P$. GCP transitions happen when the frequency of pulsar revolution matches the energy difference between two atomic states. After expanding the tidal perturbation into multipole moments and calculating its matrix elements between atomic states, one obtains the following selection rule for a process $|nlm\rangle\to |n'l'm'\rangle$ \citep{baumann2019probing}:
\begin{equation}
	\begin{aligned}
		&-m'+m_*+m=0\\
		&l+l_*+l'=2p,\text{ for } p\in\mathbb{Z}\\
		&|l-l'|\leqslant l_*\leqslant l+l'~.
	\end{aligned}\label{SelectionRules}
\end{equation}
For large circular equatorial orbits, the resonance period is
\begin{equation}
P_r=2\pi\left|\frac{\Delta m}{\Delta E}\right|~, \label{Pr}
\end{equation}
with $\Delta E\equiv E_{n'l'm'}-E_{nlm},~\Delta m\equiv m'-m$. We see that a larger $\Delta E$ generically corresponds to shorter orbital periods. 

The atomic transition produces backreaction to the binary, resulting in floating orbit and sinking orbit. We call transition with $\Delta E<0$ floating orbit, where the boson cloud loses its energy, delaying the orbital decay. In contrast, $\Delta E>0$ gives sinking orbit, where the period of binary system decreases faster with their energy given to the cloud. The total transition time is $\Delta t_{\text{tot}}=\Delta t+\Delta t_c$, where $\Delta t$ is the transition time without backreaction, and $\Delta t_c$ is the extra time caused by the backreaction. Here $\Delta t_c$ is positive for floating orbits and is negative for sinking orbits. The detailed expressions of $\Delta t$ and $\Delta t_c$ can be found in \citep{baumann2020gravitational}. A GCP transition can sometimes turn a superradiant state into a non-superradiant state with negative $\Gamma_{n'l'm'}$, where the cloud is absorbed into the BH. Interestingly, it is recently pointed out this may be avoided under certain conditions \citep{Takahashi:2021eso}.

\section{Bohr transitions vs fine and hyperfine transitions}\label{BohrVsFine}
For Bohr transitions among the lowest a few states, the energy difference is typically large, hence a short binary period. For instance, a Bohr transition from $n=3$ to $n=2$ gives a resonance period
\begin{equation}
	P_{3\to 2}=\frac{288 G M_B}{5\alpha^3}= (2.6\text{ s})\times\left(\frac{M_B}{5M_{\odot}}\right)\left(\frac{\alpha}{0.12}\right)^{-3}~.
\end{equation}
Denoting $q\equiv\frac{M_P}{M_B}$, the time left until the merger is then
\begin{equation}
	T^{(\text{merger})}_{3\to 2}=(1 \text{ day})\times\frac{(1+q)^{1/3}}{q}\left(\frac{M_B}{5M_\odot}\right)\left(\frac{\alpha}{0.12}\right)^{-8}~,
\end{equation}
suggesting that the PSR-BH binary is near the end of the inspiral phase. Although the orbital decay at this stage is significant for both pulsar timing and GW detectors, the likelihood of encountering a binary at this stage is smaller than that in the middle of the inspiral phase by at least several orders of magnitude. Yet the total number of observable PSR-BH binaries in our Galaxy is estimated to be $\mathcal{O}(10^2\text{-}10^3)$ \citep{Faucher_Gigu_re_2011,Shao:2018qpt,Chattopadhyay:2020lff}. Therefore, the event rate for Bohr transitions may be extremely low.

Another problem for Bohr transitions comes from the validity of multipole expansion. The binary separation for a typical $n=3$ to $n=2$ Bohr transition is
\begin{equation}
	R_{3\to 2}=\left(4.8\times 10^{3}\text{ km}\right)\times(1+q)^{1/3}\left(\frac{M_B}{5M_\odot}\right)\left(\frac{\alpha}{0.12}\right)^{-2}~.
\end{equation}
The size of the bosonic cloud for the state with principal quantum number $n$ is $r_n=n^2 r_1$, with $r_1=M_B\alpha^{-2}$ being the Bohr radius. For $n=3$, we have
\begin{equation}
	r_3=9M_B\alpha^{-2}=\left(4.6\times 10^{3}\text{ km}\right)\left(\frac{M_B}{5M_\odot}\right)\left(\frac{\alpha}{0.12}\right)^{-2}~.
\end{equation}
Hence the multipole expansion, in particular, the narrow resonance approximation of the $l_*=m_*=2$ quadrupole moment \citep{baumann2020gravitational} may be questionable for $q\sim 1$. This is because when $R_{3\to 2}$ is close to $r_3$, the pulsar is already moving inside the cloud, and higher multipole moments that mediate other transitions with $|\Delta m|>2$ are non-negligible. In addition, the dynamical friction of the cloud \citep{1907.13582}, upscattering effects \citep{Wong:2020qom} and the formation of molecular states \citep{2010.00008} can also have important impacts on the transition. As a result, an accurate account for Bohr transitions may require a non-perturbative treatment.

For fine ($\Delta n=0,\Delta l\neq 0$) and hyperfine ($\Delta n=\Delta l=0,\Delta m\neq 0$) transitions, however, both difficulties can be evaded. From (\ref{E}), we see that the energy difference in a fine (hyperfine) transition is smaller by a factor of $\alpha^2$ ($\alpha^3$) than a Bohr transition, leading to a much longer resonance period. This means they can happen for PSR-BH binaries in the middle phase of inspiral, which is statistically more favored. The binary separation is also enlarged by a factor of $\alpha^{-4/3}$ ($\alpha^{-2}$) for a fine (hyperfine) transition, making the multipole expansion well-defined and the $|\Delta m|=m_*=2$ approximation accurate.

Moreover, for $n\leqslant 3$, fine/hyperfine transitions always occur with $\Delta E<0$. This is because the initial cloud state has maximal helicity and the highest energy within the same $l$-multiplet, and the excitation to a higher $l$-multiplet is forbidden for $n\leqslant 3$. This suggests that these phenomenologically interesting fine/hyperfine transitions give rise to floating orbits. Thus the time spent on the GCP resonance is extended by $\Delta t_c$, further increasing the likelihood of detection. If $\Delta t_c\gg \Delta t\gtrsim T^{(\text{deplete})}$, once the binary hits the resonance band, the pulsar will be stuck on the floating orbit until the cloud depletes, which typically takes $10^8$ yr. This greatly enhances the detection likelihood.

\begin{table*}[hbt!]
	\centering 
	\caption{A comparison between Bohr transitions and fine/hyperfine transitions}\label{BVFtable}
	\begin{tabular}{cccccccccccc}
		\toprule  
		&  & \textbf{Transition} & $P_r$ (hr) & $\Delta t$ (yr) & $\Delta t_c$ (yr) & $r_n/R_r$ & $T^{(\text{growth})}$ (yr) & $T^{(\text{deplete})}$ (yr) & $T^{(\text{merge})}$ (yr) \\
		\midrule  
		&\multirow{4}{*}{\textbf{Bohr}} & $|322\rangle\to|200\rangle$ & $6.4\times10^{-4}$ & $2.8\times10^{-3}$ & $2.2\times10^{-3}$ & $0.96$ & $9600$ & $10^{13}$ & $7.9\times10^{-3}$\\
		& & $|322\rangle\to|100\rangle$ & $9.9\times10^{-5}$ & $1.9\times10^{-5}$ & $2.8\times10^{-5}$ & 3.3 & 9570 & $10^{13}$ & $5.5\times10^{-5}$\\
		& & $|311\rangle\to|21-1\rangle$ & $7.0\times10^{-4}$ & $3.6\times10^{-3}$ & $1.4\times10^{-2}$	& 0.89 & $4.7\times10^{-2}$	 & $10^{5}$ & $1.0\times10^{-2}$\\ 
		& & $|211\rangle\to|31-1\rangle$ &  $7.1\times10^{-4}$ & $3.6\times10^{-3}$ & $-1.4\times10^{-2}$ & $0.89$ & $1.7\times10^{-2}$ & $10^5$ & $1.0\times10^{-2}$\\
		\midrule
		&\multirow{1}{*}{\textbf{Fine}} & $|322\rangle\to|300\rangle$ & $1.9\times10^{-2}$ & $25$ & $6.3$ & $9.8\times10^{-2}$ & $9600$ & $10^{13}$ & $72$ \\
		\midrule
		&\multirow{4}{*}{\textbf{Hyperfine}} & $|322\rangle\to|320\rangle$& 12 & $7.5\times10^8$&$2.2\times10^7$& $1.3\times10^{-3}$&$9600$& $10^{13}$ & $2.1\times10^9$\\
		&      & $|321\rangle\to|32-1\rangle$ & $6.4$ & $1.3\times10^8$ & $2.4\times10^7$ & $2.0\times10^{-3}$ & $6.4\times10^5$ & $10^{5}$-$10^{13}$ & $3.8\times10^8$ \\
		&      & $|311\rangle\to|31-1\rangle$ & $1.3$& $1.8\times10^6$ & $5.6\times10^5$& $6.0\times10^{-3}$ & $4.7\times10^{-2}$ & $10^5$& $5.1\times10^6$ \\
		&      & $|211\rangle\to|21-1\rangle$ & 0.38 & $7.0\times10^4$ & $3.3\times10^4$ & $6.0\times10^{-3}$ & $1.7\times10^{-2}$ & $10^5$ & $2.0\times10^5$\\
		\bottomrule 
	\end{tabular} 
\end{table*}

A quantitative comparison of Bohr transitions and fine/hyperfine transitions is shown in TABLE.~\ref{BVFtable}, where the mass parameters are fixed to be $\alpha=0.12, M_B=5M_\odot$ and $M_P=1.4M_\odot$. Here we have enumerated all GCP transitions that involve states with $n,n'\leqslant 3$, and that are mediated by the $l_*=m_*=2$ quadrupole moment. Because of the uncertainty in the $T^{(\text{deplete})}$ formula (\ref{DepleteTimeApprox}), we have only kept its order of magnitude. Also note that the depletion time for the state $|321\rangle$ has not yet been computed in the literature to our best knowledge. Therefore, we only give its possible range estimated by $T_{322}^{(\text{deplete})}$ and $T_{311}^{(\text{deplete})}$. It is clear from TABLE.~\ref{BVFtable} that fine/hyperfine transitions solve all issues aforementioned, by having a much larger $P_r$, a much longer $\Delta t,T^{(\text{merger})}$, and a much smaller ratio $r_n/R_r$.

Going beyond the lowest a few states, we can find more interesting structures emerging. Given the $l_*=m_*=|\Delta m|=2$ constraint and the selection rules (\ref{SelectionRules}), one can find all possible quadrupole-mediated GCP transitions with $n,n'\leqslant n_{\text{max}}$. The formula for the total number of allowed transitions is
\begin{equation}
	N_{\text{tot},n_{\text{max}}}^{(B/F/HF)}=\frac{1}{2} \left(n_{\max }-2\right) n_{\max } \left(n_{\max }^2-2 n_{\max }+3\right)~,
\end{equation}
where $n_{\text{max}}\geqslant 4$. Within the $n$-th energy level, the number of allowed fine transitions and hyperfine transitions are given by
\begin{align}
	\nonumber N_{n}^{(F)}&=2 n^2-10 n+14\\
	N_{n}^{(HF)}&=n^2-3 n+3~,~\text{with }n\geqslant 4~.
\end{align}
We have enumerated all allowed transitions with $n_{\text{max}}=4$ and $n_{\text{max}}=8$ in FIG.~\ref{level48graph}. The transition graph with $n_{\text{max}}\geqslant 4$ neatly factorize into the direct product of four connected subgraphs. This can be understood as the consequence of the quadrupole approximation. Because $\Delta m=2$, states with odd $m$ cannot jump to states with even $m'$, vice versa. In addition, parity conservation further divides these two sets into parity-odd (odd $l$ and $l'$) families and parity-even (even $l$ and $l'$) families, leading to the four disjoint sectors. Note that although higher multipole moments ($l_*\geqslant 3$) are able to mediate Bohr transitions between these sectors, they are in general too small to influence fine/hyperfine transitions.

\begin{figure}[h!]
	\centering
	\includegraphics[width=8cm]{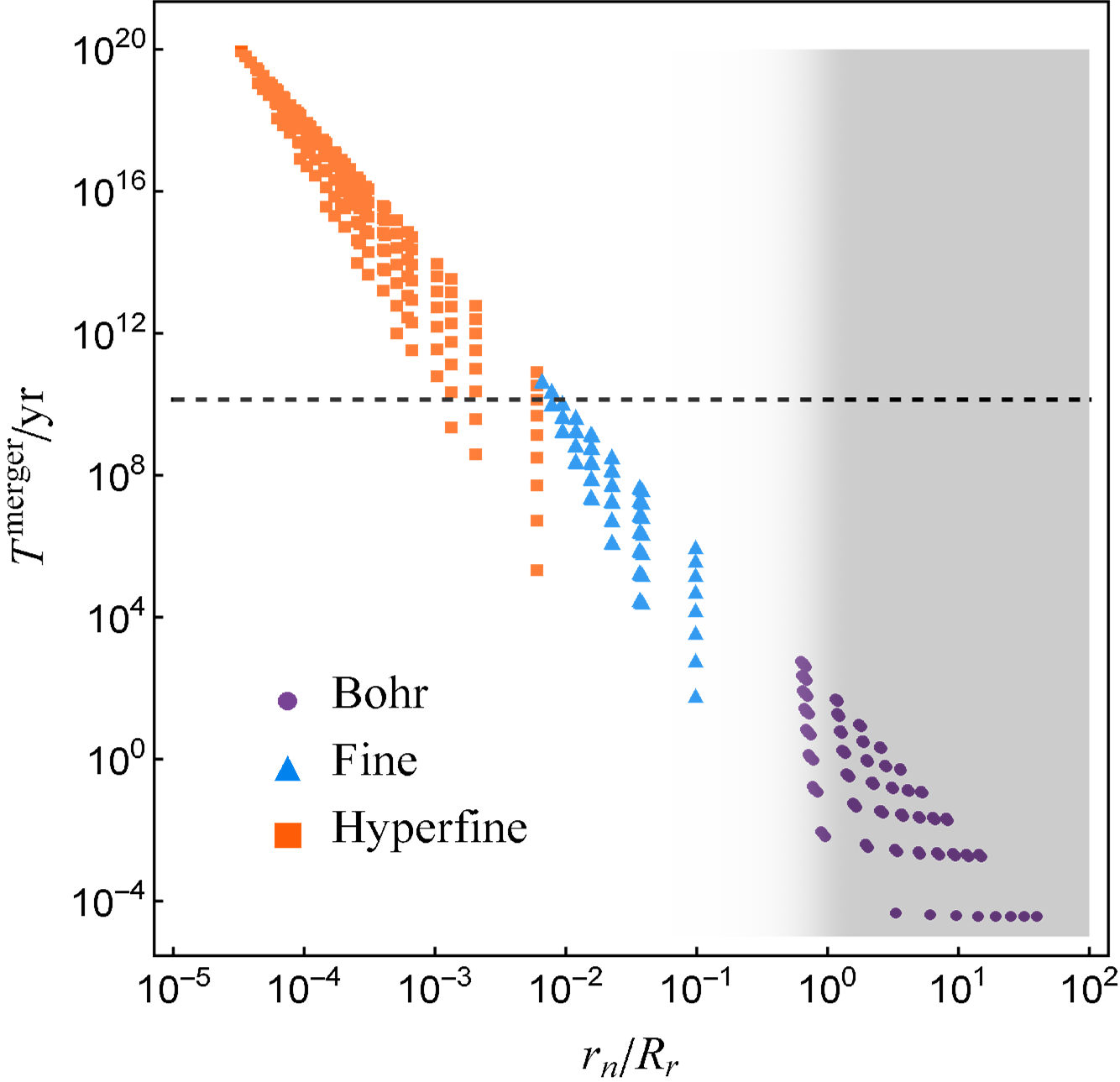}
	\caption{The distribution of Bohr/fine/hyperfine transitions in the parameter space spanned by the time $T^{\text{merger}}$ left until merger as well as the ratio of cloud size and binary separation $r_n/R_r$. Here we have considered all quadrupole-mediated GCP transitions up to $n_{\text{max}}=10$. There are in total $N_{\text{tot},10}^{(B/F/HF)}=3320$ transitions. The parameters are again chosen as $\alpha=0.12$, $M_B=5 M_\odot$ and $M_P=1.4 M_\odot$. The gray region is where the multipole expansion and narrow resonance approximation break down. The dashed line represents the age of the universe. A smaller ratio $r_n/R_r$ gives better quadrupole approximation and a longer $T^{\text{merger}}$ enhances the event rate\footnote{Notice that a $T^{\text{merger}}$ longer than the age of the universe does not mean the transition cannot happen, nor the binary does not exist. It just means that the binary is going to spend an extremely long time in the inspiral phase.}. Hence the advantage of fine/hyperfine transitions over Bohr transitions is obvious.}
	\label{BvFvHF}
\end{figure}

The advantage of fine/hyperfine transitions over Bohr transitions persists as more GCP transitions are included. In FIG.~\ref{BvFvHF}, we have shown the time left until merger as well as the ratio of cloud size and binary separation for all transitions up to $n_{\text{max}}=10$. It is clear that the three types of transitions occupy different regions in the parameter space. Most Bohr transitions lie in the gray region where the multipole expansion breaks down, while fine/hyperfine transitions are far safer. Fine/hyperfine transitions also correspond to a much longer time before merger, hence a higher event rate.

\section{Uncovering fine and hyperfine structures: Pulsar timing accuracy}\label{FineHyperfine}
The direct observation of GCP transitions for a PSR-BH binary relies on an accurate measurement of orbital motion, which is recorded as modulations in the Rømer delay of pulsar time-of-arrivals. Unlike Bohr transitions, fine/hyperfine transitions occur at a much longer orbital period, where the GW emission is still weak. The resonance frequency lies in the range of the space-based GW detectors such as LISA. At a low orbital frequency, the corresponding orbital decay is much slower. Such a weak effect may require a long-term observation that lasts more than a decade. As a comparison, LISA only has a lifetime of 4-6 years \citep{LISA}. Therefore, it is questionable whether the first generation of space GW detectors \citep{LISA,Taiji,tianqin,decigo} are precise enough to probe the GCP transitions in due time. In contrast, radio telescopes are earth-based and can last many decades. For instance, the Arecibo telescope built in 1963 had been functioning for 57 years before its tragic collapse in 2020. Thus a long-term observation of a PSR-BH binary may reveal the tiny deviations of orbital decay and uncover the fine/hyperfine structure of the gravitational atom.

The orbital decay of the PSR-BH binary, according to \citet{Hulse} and \citet{Hulse2}, can be observed by recording the periastron time shift
\begin{equation}
	\Delta_P=t-P(0)\int^t_0\frac{1}{P(t)}{\rm{d}}t'~.\label{PerishiftDef}
\end{equation}
If we consider a small timescale with respect to that of a significant orbital decay, we can linearize the period change by $P(t)\simeq P(0)+\dot{P}t$. The periastron time shift then increases quadratically with observation time:
\begin{equation}
	\Delta_P=\frac12\frac{\dot{P}}{P}t^2.
\end{equation}
Under the influence of isolated atomic transitions, the orbital decay due to GW emission can be approximated as \citep{oringin}
\begin{equation}
	\begin{aligned}
		\dot{P}&=-\frac{96}{5}(2\pi)^{8/3}(GM_B)^{5/3}\frac{q}{(1+q)^{1/3}}P^{-5/3}\\
		&\times\frac{1}{1\pm\frac{\Delta t_c}{\Delta t}\times\Pi(\frac{P-P_r}{\Delta P_r})},\label{Pdot}
	\end{aligned}
\end{equation}
where $\Pi$ is the Heaviside-Pi window function that characterize the bandwidth of the GCP resonance, $i.e.$, $\Delta P_r\simeq\frac{0.6q}{1+q}P_r$. The deviation of $P(t)$ and hence $\Delta_P(t)$ from the general relativity prediction will be the signal that we are after.

In order to successfully detect GCP resonances, we need to ensure that the uncertainty $\sigma_{\Delta_P}$ is smaller than total Periastron time shift $\Delta_P$. Notice that the periastron time shift is not measured from its definition (\ref{PerishiftDef}), which demands an accurate measurement of the orbital period as a function of time. Rather, we measure the periastron time shift directly by counting the accumulated orbital periods. We determine one orbital period by counting the number of pulses between two pulses with equal Rømer delay and equal time derivative of Rømer delay. In order to resolve an orbital period, the maximal Rømer delay must be larger than the pulse width $w$, which is on a similar order as the pulse period $\tau$. Therefore, we demand
\begin{equation}
	\frac{2R_r}{1+q}>\tau~,\label{RomerDelayConstraint}
\end{equation}
which automatically implies $P_r>\tau$. The error during one continuous observation window is also $w\sim\tau$. Suppose we can observe the pulsar for $t_{obs}$ every day, which means we can measure $t_{obs}/P$ periods every day. That is, for every single continuous measurement, the error can be estimated by $\frac{\tau}{\text{min}(t_{obs},t)/P}$. If we observe for $0<t\leqslant T_{obs}$, where $T_{obs}$ is the maximal observation time, then the number of independent measurement is $\lceil t/1 \text{ day}\rceil$, where $\lceil\rceil$ is ceil function. In summary, the uncertainty for Periastron time shift is 
\begin{equation}
	\sigma_{\Delta_P}=\frac{1}{\sqrt{\lceil t/1 \text{day}\rceil}}\frac{\tau}{\text{min}(t_{obs},t)/P}~. \label{PeriUncertainty}
\end{equation}
We can detect the GCP transitions only if the difference between the periastron time shift with transition backreaction and that without it is greater than the observation uncertainty,
\begin{equation}
	\Big|\Delta_P|_{GCP}-\Delta_P|_{GR}\Big|>\sigma_{\Delta_P}~.\label{TimingAccuracyConstraint}
\end{equation}
In addition, there are two more constraints on the model parameters. Namely, the superradiant time scale of boson cloud should be short enough to observe, and the cloud should be stable on an astrophysical timescale \citep{baumann2020gravitational},
\begin{equation}
T^{\text{(growth)}}\lesssim10^6\text{ yrs,  }T^{\text{(deplete)}}\gtrsim10^8\text{ yrs}~.\label{CloudConstraint}
\end{equation}

\begin{figure*}[htbp]
	\centering
	
	\subfigure{
		\begin{minipage}[t]{0.45\linewidth}
			\centering
			\includegraphics[width=7.5cm]{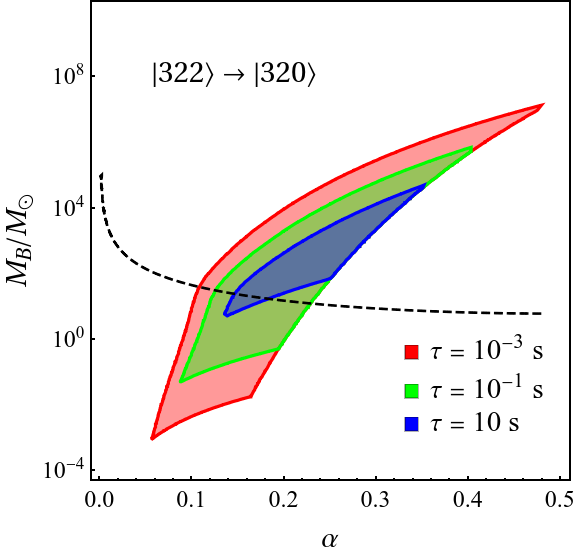}
		\end{minipage}
	}
	\subfigure{
		\begin{minipage}[t]{0.45\linewidth}
			\centering
			\includegraphics[width=7.5cm]{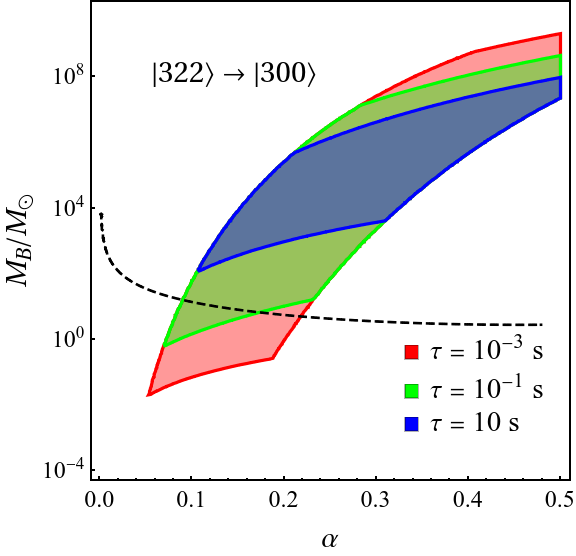}
		\end{minipage}
	}
	
	\subfigure{
		\begin{minipage}[t]{0.45\linewidth}
			\centering
			\includegraphics[width=7.5cm]{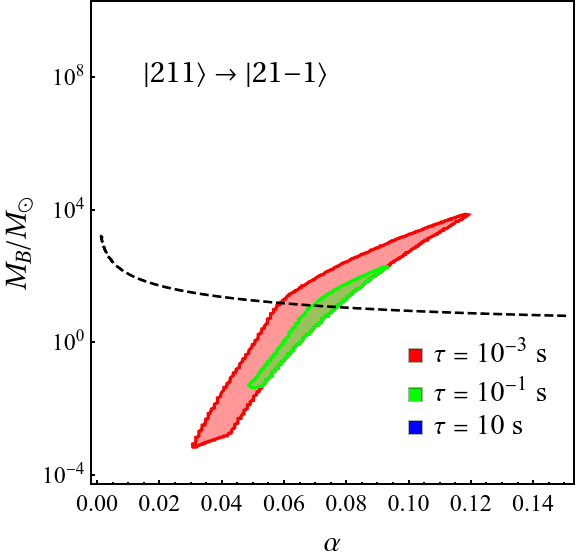}
		\end{minipage}
	}
	\subfigure{
		\begin{minipage}[t]{0.45\linewidth}
			\centering
			\includegraphics[width=7.5cm]{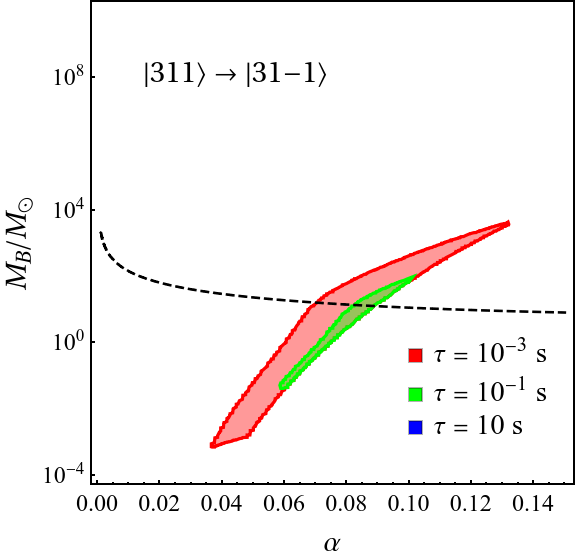}
		\end{minipage}
	}
	\subfigure{
		\begin{minipage}[t]{0.45\linewidth}
			\centering
			\includegraphics[width=7.5cm]{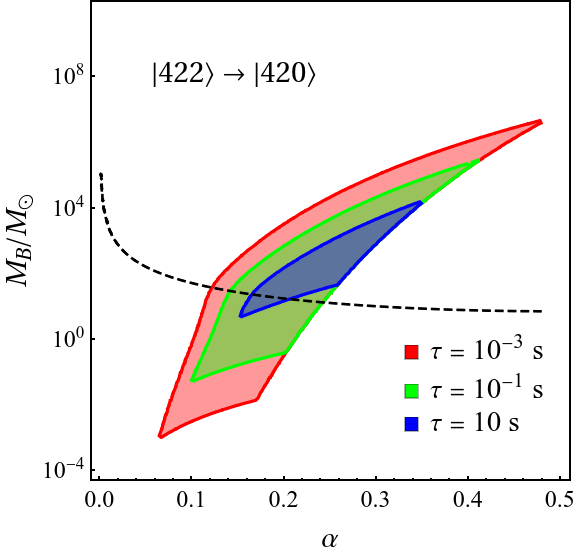}
		\end{minipage}
	}
	\subfigure{
		\begin{minipage}[t]{0.45\linewidth}
			\centering
			\includegraphics[width=7.5cm]{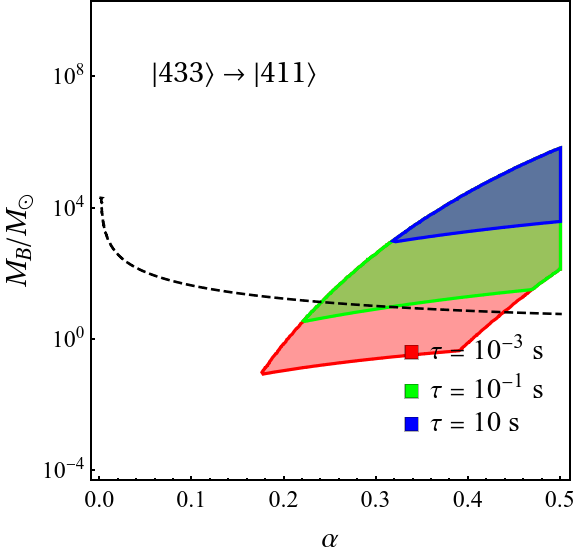}
		\end{minipage}
	}
	
	\centering
	\caption{The feasible regions for $\alpha$, $M_B$ and different pulsar periods $\tau$. Here we have taken the mass of the pulsar to be $M_P=1.4M_{\odot}$ and assumed an observation time $T_{obs}=1$ decade. The black dashed line given by $\Delta t=\Delta t_c$ divides the parameter space into the upper half with strong backreaction and the lower half with weak backreaction. The timing accuracy edge acquires a turning point near this line because the periastron time shift has different asymptotic behaviors across the line.}
	\label{FeasibleRegions}
\end{figure*}

Combining the constraints (\ref{RomerDelayConstraint}), (\ref{TimingAccuracyConstraint}) and (\ref{CloudConstraint}), we obtain the feasible parameter region for the fine/hyperfine transitions\footnote{Note that the transition $|321\rangle\to |32-1\rangle$ is not shown due to our lack of information on $T_{321}^{(\text{deplete})}$.} shown in FIG.~\ref{FeasibleRegions}. Overall, transitions starting with $|322\rangle$ give a wide range of parameter region that covers $10^{-3}M_\odot<M_B<10^8 M_\odot$ and $0.06<\alpha<0.5$. In contrast, transitions starting with $|311\rangle$ and $|211\rangle$ allow a relatively limited parameter space with significantly smaller $\alpha$. Notice that the right edge of the parameter space is constrained by the depletion time, which is subjected to an $\mathcal{O}(10)$ uncertainty. The lower edge is constrained by the Rømer delay resolvability (\ref{RomerDelayConstraint}). The upper edge is constrained by the timing accuracy of periastron time deviation (\ref{TimingAccuracyConstraint}). The left edge is either constrained by the timing accuracy (for hyperfine transitions) or the cloud growth time (for fine transitions). 

\begin{figure}
\centering
\includegraphics[width=8cm]{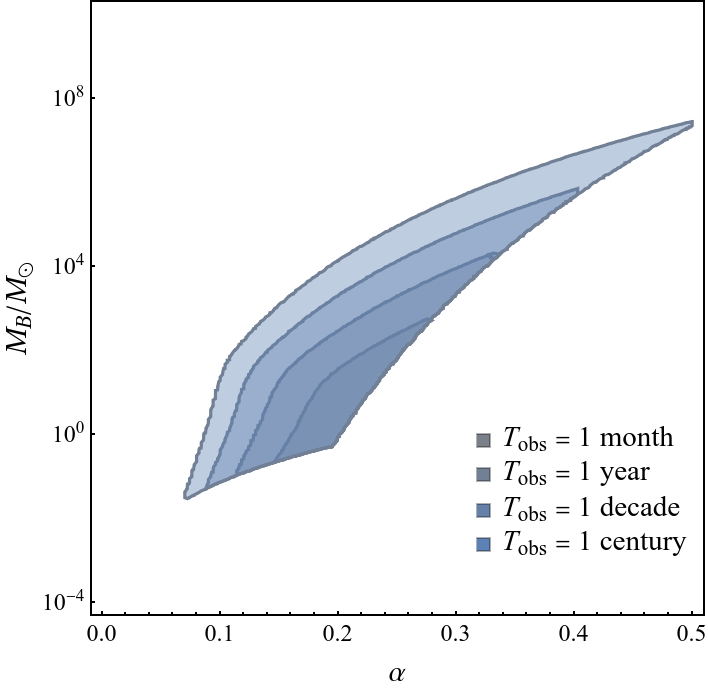}
\caption{The feasible parameter space for $\alpha$, $M_B$ and different observe time $T_{\text{obs}}$ for the hyperfine transition $|322\rangle\to|320\rangle$. Here we have chosen $M_P=1.4M_{\odot}$ and $\tau=0.1$s.}
\label{Fig2}
\end{figure}

\begin{figure*}[!htpb]
	\centering  
	\subfigure[Allowed GCP transitions up to $n_\text{max}=4$]{
		\begin{minipage}[t]{\linewidth}
			\centering
			\includegraphics[width=10cm]{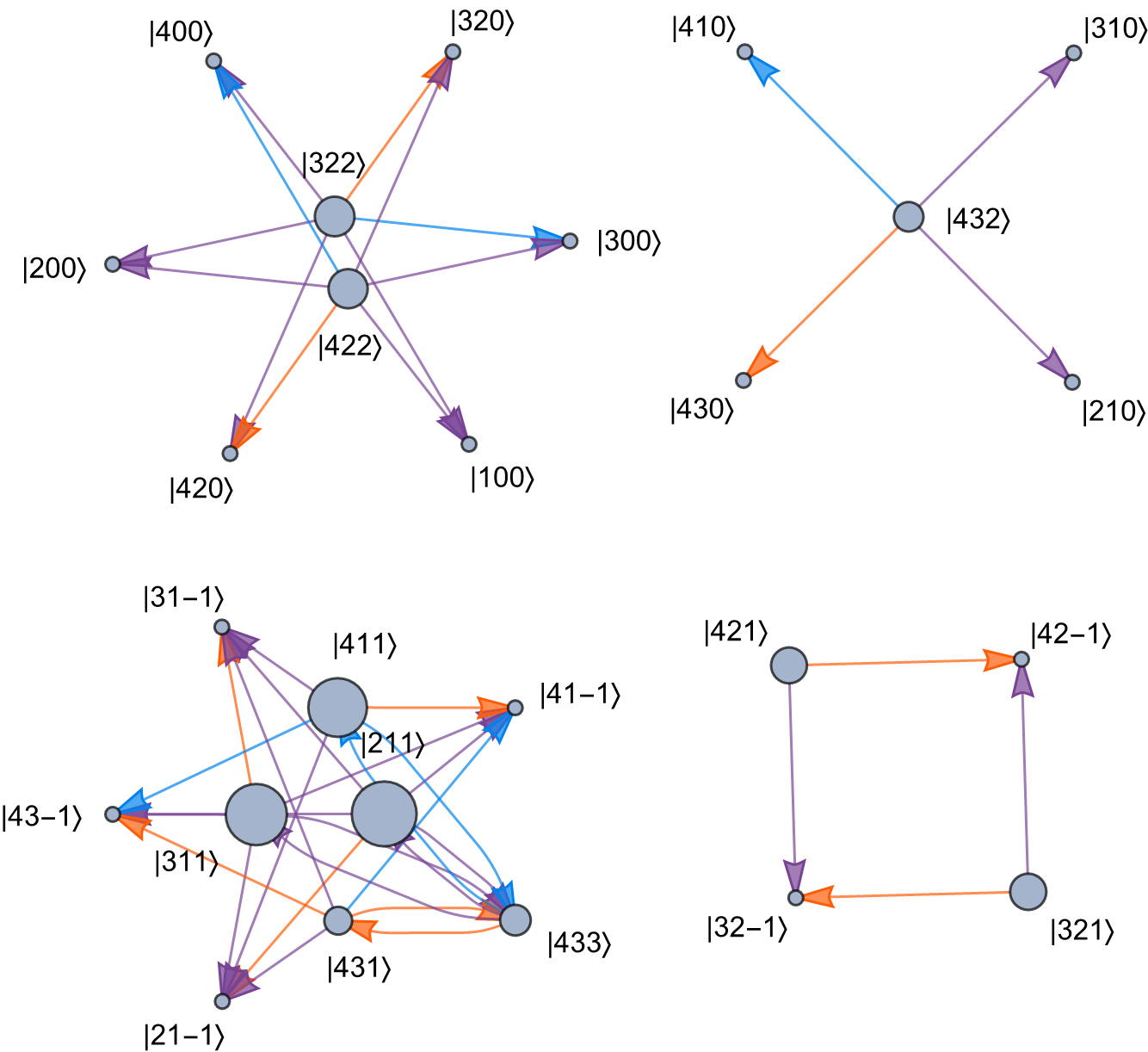}
		\end{minipage}
	}  
	
	\subfigure[Allowed GCP transitions up to $n_\text{max}=8$]{
		\begin{minipage}[t]{\linewidth}
			\centering
			\includegraphics[width=15cm]{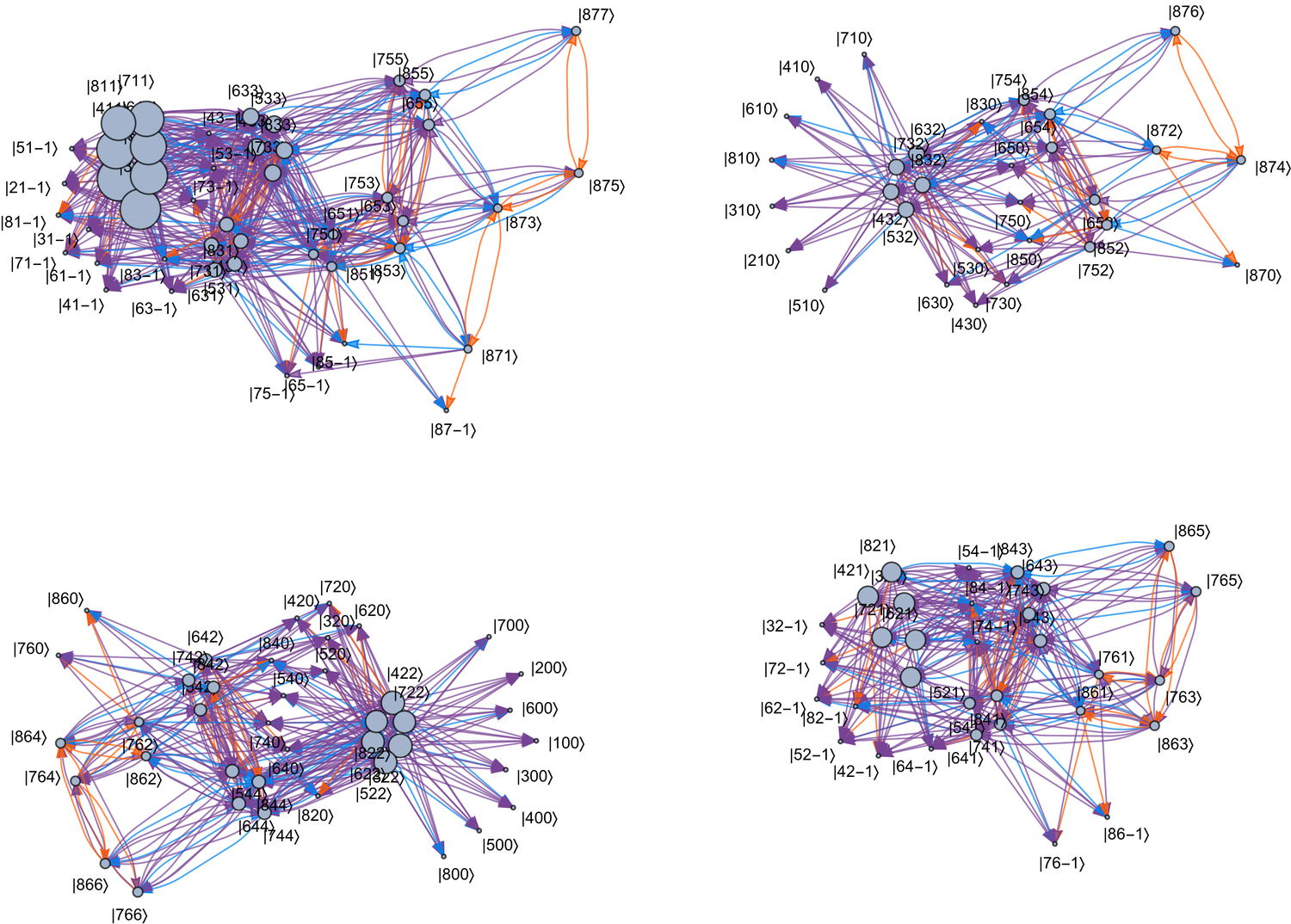}
		\end{minipage}
	} 
	\caption{The transition graph for states with $n,n'\leqslant n_{\text{max}}$. The purple, blue and orange arrows represent Bohr transitions, fine transitions and hyperfine transitions, respectively. The size of the vertices indicates the superradiance growth rate, $i.e.$, a large vertex corresponds to a fast-growing cloud.} 
	\label{level48graph}
\end{figure*}

Thus we see that timing accuracy plays an important role in probing hyperfine transitions. Naturally, pulsars with shorter rotations period $\tau$ provide a finer resolution of the orbital motion, thereby increasing the timing accuracy. Alternatively, for a given PSR-BH system, one can also extend the observation time $T_{\text{obs}}$ to increase the timing accuracy. This fact is demonstrated in FIG.~\ref{Fig2} for the hyperfine transition $|322\rangle\to |320\rangle$. Some parameter choices may require decades of observation for a clear detection, a task suitable only for ground-based apparatus such as radio telescopes.

The fine/hyperfine transitions for states with $n\geqslant4$ are qualitatively similar. Since the superradiance growth rate and the GW depletion rate is not very sensitive to the principal quantum number, the feasible parameter regions for $|n22\rangle\to |n20\rangle$ and $|n22\rangle\to |n00\rangle$ are similar to that plotted in FIG.~\ref{FeasibleRegions}. Transitions starting with higher angular quantum number $l$ generally require a larger $\alpha$, since their growth rate are further suppressed by powers of $\alpha$.

\section{Conclusion}\label{Conclusion}
In this paper, we focused on probing fine/hyperfine GCP transitions in a PSR-BH binary with pulsar timing. Starting from a general review of the gravitational atom and GCP, we pointed out the problems faced by Bohr transitions. Namely, the Bohr transitions may be extremely rare because they happen near the end of the inspiral phase. The quadrupole narrow resonance approximation may also become invalid when the binary separation is comparable to the cloud size. Then we showed that these problems can be evaded in the fine/hyperfine transitions, which typically enjoy a longer resonance period and a greater binary separation. All fine/hyperfine transitions with $n\leqslant 3$ lead to floating orbits that delay the orbital decay, further enhancing the likelihood of observation. The advantage persists for higher energy levels, as the three types of transitions occupy distinct regions in the parameter space. The subsequent analysis of pulsar timing accuracy demonstrates the feasibility of detecting fine/hyperfine transitions in GCP. In particular, we find that the fine transition $|n22\rangle\to |n00\rangle$ and hyperfine transitions $|n22\rangle\to|n20\rangle$ give wide parameter regions that can be probed via pulsar timing. Increasing the total observation time also increases the timing accuracy, making the detection of transitions with large black hole mass possible. In the spirit of multi-messenger astronomy, due to the long lifetime of radio telescopes and the stringent accuracy requirement, the observation of fine/hyperfine transitions via this PSR-BH-radio channel should serve as a complement to observing Bohr transitions via the BH-BH-GW channel.

However, there are still many questions left unanswered in the current work, and we hope to address them in the future. We conclude this paper by mentioning a few of them. First, our estimation of boson cloud depletion time is based on an approximate formula with a considerable amount of error, especially at large $\alpha$ and when $l\neq m$. We hope to improve our constraints in the future using more accurate numerical solutions of cloud depletion. Second, we have argued that the event rate of fine/hyperfine transitions is greatly enhanced by their long orbital periods and floating orbits, yet we did not give any explicit estimation of the likelihood of observation. To perform such an analysis, one needs to consider the initial distribution of BHs and pulsars in the Galaxy as well its evolution history. Third, in addition to pulsar timing, the measurement of the Doppler effect and the ellipsoidal modulation of a White Dwarf (WD) in a binary is nowadays accurate enough for detecting the orbital decay \citep{1,2,3}. Since there are more WDs than pulsars in the Galaxy, it is interesting to consider probing GCP resonances in WD-BH binaries, whose event rate can be further enhanced. We can also consider the Doppler effect of a generic Neutron Star (NS) which may not necessarily be a pulsar. This possibility becomes more exciting after the recent LIGO-Virgo discovery of two NS-BH mergers by \citet{LIGOScientific:2021qlt}. The question is, of course, whether the accuracy can reach the requirement of detecting the \textit{deviations} in the orbital decay due to GCP resonances. We leave a detailed analysis to future works.

\section*{Acknowledgment} 
We would like to thank Qianhang Ding and Chon Man Sou for helpful discussions. This work was supported in part by the CRF grant C6017-20GF by the RGC of Hong Kong SAR, and the NSFC Excellent Young Scientist (EYS) Scheme (Hong Kong and Macau) Grant No. 12022516.\\\\\\\\\\\\\\\\\\\\\\\\\\

\nocite{*}
\bibliographystyle{apalike}
\bibliography{reference}

\end{document}